\documentclass[conference,compsoc]{IEEEtran}
\usepackage{graphicx}
\usepackage{booktabs}
\usepackage{amsmath}
\usepackage{multirow}
\usepackage{xcolor}
\usepackage[hidelinks]{hyperref}
\usepackage{url}
\usepackage{tcolorbox}
\newtcolorbox{answerBox}[1][]{
    colback=gray!25,
    colframe=gray!20,
    boxrule=0.2mm,
    arc=0.5mm,
    width=\columnwidth,
    left=0.2mm,
    right=0.2mm,
    top=0.2mm,
    bottom=0.2mm,
    #1}

\newcommand{\parhead}[1]{\par\vspace{1.5pt}\noindent\textbf{#1}\enspace}

\title{A Footprint-Sensitive Cross-GPU Timing Channel\texorpdfstring{\\}{ }Through the Shared Host Memory Subsystem}

\author{\IEEEauthorblockN{Wei Song\textsuperscript{1}, Yuxin Cao\textsuperscript{2}, Christy Jie Liang\textsuperscript{3}, Xi Zheng\textsuperscript{4}, Leo Yu Zhang\textsuperscript{1}, Xiao Cheng\textsuperscript{4}}
\IEEEauthorblockA{\textsuperscript{1}Griffith University, Australia\\
\textsuperscript{2}National University of Singapore, Singapore\\
\textsuperscript{3}The University of Technology Sydney, Australia\\
\textsuperscript{4}Macquarie University, Australia}}

\begin{document}
\maketitle

\begin{abstract}
Multi-GPU servers assign tenants to separate GPUs, but both GPUs' DMA data traffic and IOMMU page-table walks converge on the same host memory controller. We show that an unprivileged co-tenant on one GPU can time its own reads to pinned host memory to detect contention from a victim on another GPU. The attacker accesses only its own buffer, never the victim's page tables or data, and uses no NVLink, shared IOTLB, or hardware performance counters. The channel is footprint-sensitive: the probe-latency shift tracks the number of distinct host pages the victim touches rather than its throughput, exposing the structure of the victim's memory access rather than merely its presence. We build three attacks on this channel. A covert-channel attack transmits between colluding tenants at 20~bit/s with near-zero error. A length-inference attack recovers the per-request output length of a co-resident language model at a held-out Spearman correlation of 0.99, across five models from 0.5 to 14 billion parameters on three platforms. A refusal-inference attack reveals whether the model refused a query, a semantic property of the victim's private input, at a balanced accuracy up to 0.955. The attacks cross process, container, and Unix-user boundaries and reproduce on two GPU generations and two serving runtimes. A production vLLM+LMCache stack stays below the leakage floor when its traffic is bursty, bounding the channel to sustained device-to-host regimes. A replicated \texttt{iommu=off} control traces the mechanism to IOMMU page-walk contention in shared host DRAM, and a cross-NUMA test localizes it to the memory controller, yielding two defenses, controller-separation placement and footprint flattening, both of which reduce the attacker to chance. Our findings motivate future research on timing-channel threats under concurrent LLM serving, where the shared host infrastructure beneath isolated accelerators remains exposed.
\end{abstract}

\section{Introduction}

Multi-GPU servers place mutually distrusting tenants on physically separate GPUs. Cloud providers rent individual GPUs to different customers, schedulers pack unrelated jobs into one chassis, and inference services co-locate models whose inputs and outputs must remain private~\cite{ppautodl,medshield}. Each GPU has its own compute units, caches, and device memory, making device assignment a natural isolation boundary. Yet workloads on separate GPUs still access the same host memory subsystem. We ask whether this shared resource lets an unprivileged tenant learn about a workload on another GPU, even without NVLink or peer-to-peer access.

\begin{figure*}[t]\centering
\includegraphics[width=0.98\linewidth]{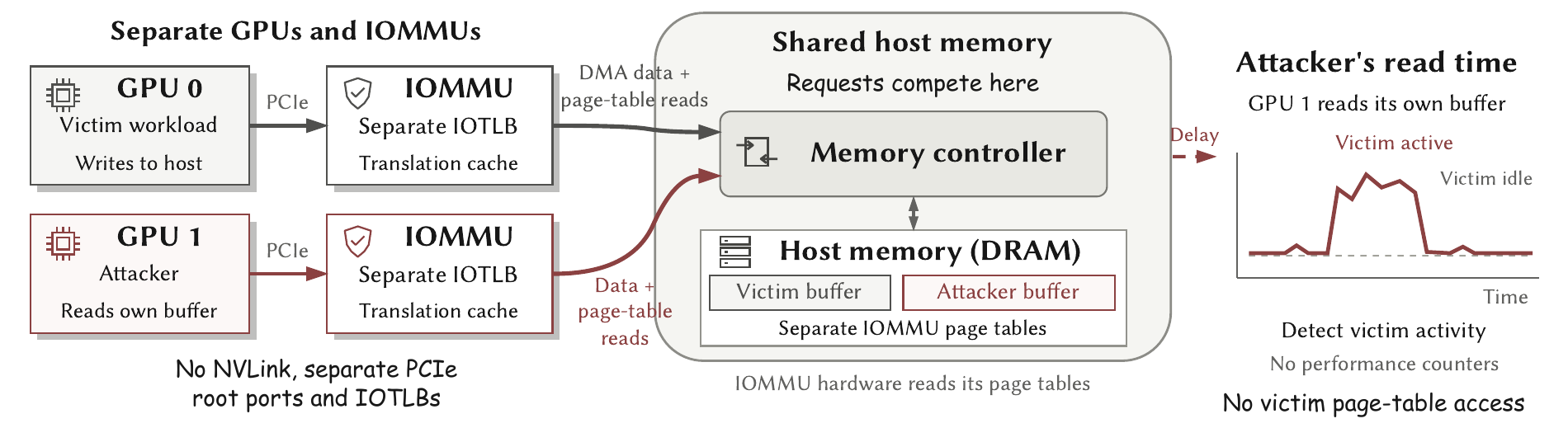}
\caption{Shared host memory and attack mechanism. GPU~1 measures the time taken to read its own pinned host buffer. Victim DMA data and IOMMU page-table reads compete for the same memory controller, which can lengthen these reads. Page-table reads are performed by IOMMU hardware. The schematic waveform compares read time during victim activity with the dashed idle baseline. The dashed arrow shows the effect of contention.}
\label{fig:overview}
\end{figure*}

Figure~\ref{fig:overview} illustrates how a GPU accesses host memory. Direct memory access (DMA) lets a GPU read or write host memory (DRAM) over PCIe without the CPU copying each byte. CUDA workloads commonly use pinned host memory: pages kept resident while the GPU accesses them~\cite{nvidiaCudaMemory}. An input/output memory management unit (IOMMU) translates device-issued DMA addresses to permitted host physical addresses using I/O page tables maintained by system software~\cite{linuxDmaHowto}. Inside each IOMMU, an I/O translation lookaside buffer (IOTLB) acts as a translation cache that stores recent address mappings. If a required mapping is not cached, the IOMMU hardware looks it up by reading its multi-level page tables. This lookup is called a page-table walk~\cite{devious,iotlbsc}. The two GPUs in our setup use separate IOMMUs and separate IOTLBs, but their I/O page tables and DMA buffers reside in host DRAM served by the same memory controller. Both DMA data traffic and IOMMU page-table reads therefore use this shared controller.

Our attack exploits the shared host memory controller. The attacker allocates its own mapped, pinned host buffer and runs a kernel on GPU~1 that repeatedly reads this buffer. It uses the GPU global timer to measure how long these reads take, and we call this measurement the attacker's \emph{read time}. When GPU~0 transfers data to host memory, the victim's DMA data traffic and the page-table reads performed by its IOMMU compete with the attacker's reads at the same controller. This competition can make the attacker's read time longer. Repeating this measurement can reveal when the victim is active and how long it remains active. The attacker accesses neither the victim's data nor its page tables and uses no hardware performance counters. The IOMMU hardware performs the page-table walks, and the attacker observes only their possible effect on its own read time.

This shared-host path extends beyond the GPU-local resources targeted by prior cache, TLB, and scheduler attacks~\cite{invalidatecompare,tunnels,tlbcrossvm}. Prior cross-GPU channels exploit multi-GPU interconnects or peer-to-peer communication~\cite{spygpubox,nvbleed}, while DevIOus and IOTLB-SC exploit contention in a shared IOTLB~\cite{devious,iotlbsc}. Our channel requires neither an inter-GPU link nor a shared translation cache, because the coupling occurs where the otherwise separate device paths of the two GPUs reach shared host memory.

The signal has a distinctive property: it is footprint-sensitive. At a fixed DMA rate of 4~GB/s, a victim that reuses a small host-memory region leaves the attacker's read time near its idle baseline, while the same rate spread over a large region produces a strong separation. The increase in read time grows with the number of distinct pages touched and saturates past a knee near 24~MiB. Bandwidth also affects the signal, but bandwidth alone does not explain it. This dependence is consistent with translation pressure: touching more pages can exceed IOTLB coverage and generate more page-table-walk traffic. It gives the attacker a signal about the victim's memory-access footprint as well as its active intervals.

Two controls narrow the mechanism. A replicated \texttt{iommu=off} on/off/on cycle removes the signal when translation is disabled and restores it when translation is re-enabled, supporting an IOMMU-dependent origin. The separate remapping units exclude a shared-IOTLB explanation, and together with the matched-bandwidth footprint contrast, this points to page-table-walk contention in shared host DRAM. This is a mechanism inference from controlled timing measurements, and we do not observe victim walks directly. A device-order-verified cross-NUMA test further localizes the coupling to the memory controller, since a same-controller victim increases the attacker's read time by hundreds of nanoseconds, whereas a cross-controller victim changes it by less than one nanosecond.

We build three attacks from these timing measurements. A covert-channel attack lets colluding tenants communicate at 20~bit/s with near-zero error. A length-inference attack uses the duration of generation-related contention to recover a co-resident language model's per-request output length at a held-out Spearman correlation of 0.99. A refusal-inference attack uses the same duration signal to infer whether the model refused a query, when refusals are systematically shorter than accepted responses, reaching a balanced accuracy up to 0.955. The inference attacks require no interaction with the victim's API. We evaluate five models from 0.5 to 14 billion parameters across three platforms and two serving runtimes. The underlying channel crosses process, container, and Unix-user boundaries.

Exploitability depends on the victim's transfer regime. CPU-offloaded models with sustained device-to-host traffic leak, whereas a production vLLM+LMCache stack with bursty transfers stays below the leakage floor. End-to-end inference completes on about two thirds of fresh victim processes, with failures at trace segmentation, and refusal inference requires the victim model to produce systematically shorter refusals, which not all models do. The mechanism also suggests two defenses. Placing tenants on separate memory controllers near-eliminates the signal, and footprint flattening drives an adaptive attacker to chance.

We make the following contributions:
\begin{itemize}\itemsep2pt
\item We identify a footprint-sensitive cross-GPU timing channel through shared host memory. An unprivileged attacker measures only its own GPU's accesses to its own pinned host buffer, without NVLink, a shared IOTLB, victim page-table access, or hardware performance counters.
\item We demonstrate a covert-channel attack at 20~bit/s with near-zero error, a length-inference attack with a held-out Spearman correlation of 0.99, and a refusal-inference attack with balanced accuracy up to 0.955.
\item We provide controlled evidence for IOMMU page-walk contention and memory-controller locality, and evaluate controller-separation placement and footprint flattening as defenses.
\end{itemize}

\section{Background and Threat Model}
\label{sec:bg}
We detail the translation and memory-access paths in Figure~\ref{fig:overview} before specifying the attacker's capabilities.

\parhead{DMA and Address Translation.} DMA lets a GPU access host buffers over PCIe without the CPU issuing a load or store for each transferred value. Pinned pages remain resident during device access, and mapping a pinned buffer into a CUDA address space lets a kernel access that buffer directly~\cite{nvidiaCudaMemory}. Pinning does not grant access to another tenant's memory. With IOMMU translation enabled, device DMA addresses are mapped to permitted host physical pages through I/O page tables maintained by system software~\cite{linuxDmaHowto}. The IOMMU's translation is distinct from the GPU's own virtual-address translation.

\parhead{IOTLBs and Hardware Page-Table Walks.} An IOTLB caches translations so that repeated accesses to a mapped page need not repeat the full lookup. A miss can trigger a hardware walk through the IOMMU's multi-level page tables~\cite{devious,iotlbsc}. The entries reside in host memory, although translation and memory caches can serve some lookups without a DRAM access. A DMA stream that exceeds translation-cache coverage can therefore add walk traffic to its payload traffic. The attacker neither traverses these tables in software nor measures their contents or miss counts.

\parhead{Separate Translations, Shared Memory Service.} On our workstation, GPU~0 and GPU~1 sit behind separate VT-d remapping units (\texttt{dmar5} and \texttt{dmar6}), each with its own IOTLB. Their I/O page tables and host buffers occupy separate memory regions, but both use the same host memory controller and DRAM. The attacker's probe reads only its own buffer, and increased latency can reflect competition for that shared service. Other candidate causes include shared PCIe resources, driver serialization, and platform power effects. Section~\ref{sec:mech} uses cross-agent and IOMMU-toggle controls to narrow these candidates. The toggle and footprint dependence support page-walk contention, while the cross-NUMA control in Section~\ref{sec:reach} establishes memory-controller locality. These controls neither measure walks directly nor exclude every DMA-path change from disabling the IOMMU on the Workstation.

\parhead{Threat Model.} We consider an unprivileged co-tenant assigned one GPU while the victim uses a separate GPU on the same physical host. The channel requires their host-memory traffic to share a memory controller. The two tenants run as separate processes, containers, or Unix users, and we demonstrate the underlying channel across all three boundaries. The attacker can allocate its own device memory and mapped pinned host memory, run ordinary CUDA code, and read the GPU global timer to time its own host-buffer accesses. It has no root privilege, no hardware performance counters, and no access to the victim's data, CPU or IOMMU page tables, GPU execution context, or API endpoint. It cannot inspect or modify the victim's mappings. The global timer supplies elapsed time, not a count of IOTLB misses or page walks.

The passive inference attacks require no victim cooperation or requests to its API. They aim to detect generation activity, recover per-request output length, or infer refusal from generation duration under the evaluated response-length relationship. The covert-channel attack instead assumes a colluding sender on the other GPU that deliberately modulates its DMA activity. Neither setting requires NVLink, PCIe peer-to-peer access, a shared IOTLB, MIG, vGPU, or physical access. The observable in both settings is the latency of the attacker's own GPU reads, and the attacker infers any information about the victim from contention in that latency trace alone.

\section{Channel and Attacks}
\label{sec:channel}
This section presents the cross-GPU timing channel and the attacks built on it. We first describe the measurement primitive that the attacker deploys and the footprint sensitivity that makes the channel discriminative, and then define the three attacks, distinguishing the covert channel from the two passive side-channel attacks, with each attack's prerequisites, inference procedure, and security implications. Section~\ref{sec:eval} evaluates all three.

\begin{figure*}[t]\centering
\includegraphics[width=0.98\linewidth]{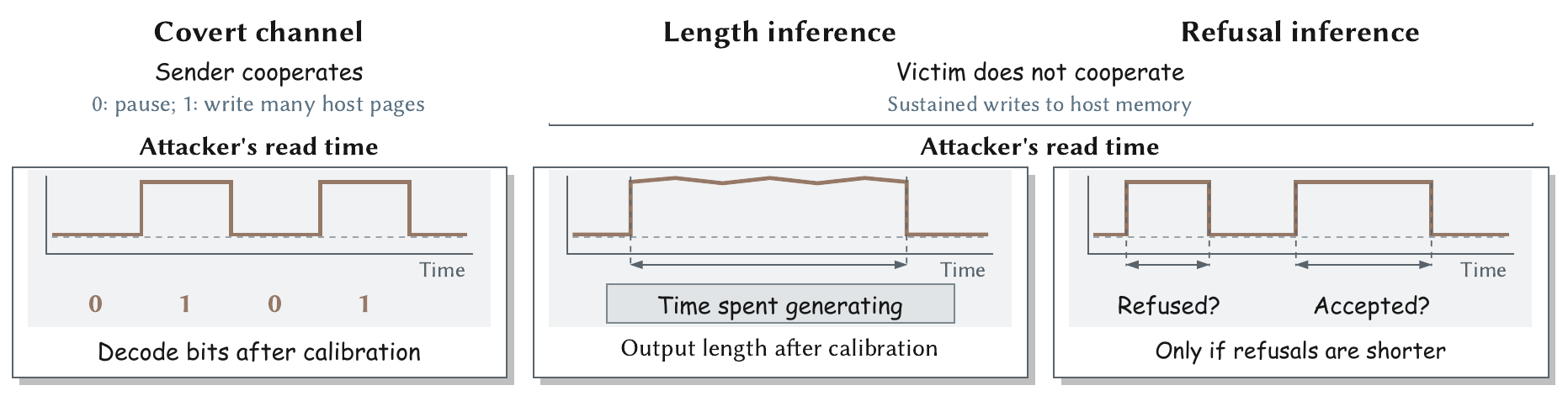}
\caption{Three attacks built on the read-time signal. Left: calibrated high/low levels encode bits. Center: generation duration estimates output length in tokens after calibration. Right: the same duration feature predicts refusals when refusals are shorter than accepted responses. The two passive attacks require sustained victim writes to host memory. Waveforms are schematic.}
\label{fig:channel-attacks}
\end{figure*}

\parhead{Primitive.} The attacker allocates a mapped, pinned host buffer and performs a single-thread pointer chase over it from a kernel on the second GPU, timing a block of chased reads with the global timer. Each read is a round trip from the GPU through the root complex to host memory, and the per-access latency rises under host-memory contention. This latency is the attacker's read time shown in Figures~\ref{fig:overview} and~\ref{fig:channel-attacks}. The primitive uses no performance counters. Because the absolute latency depends on the load of the attacker GPU itself, we compare conditions only within a run and in randomized order, as Section~\ref{sec:eval} details.

\parhead{Footprint Sensitivity.} Using this primitive, we find that the channel is footprint-sensitive: the probe-latency shift scales with the number of distinct host-memory pages the victim's DMA touches, not with raw throughput. A large-footprint transfer that spans many pages drives a strong signal, while a small-footprint transfer at the same bandwidth does not. Section~\ref{sec:mech} establishes this with a dose-response sweep and traces it to IOMMU page-walk contention in shared host DRAM. The property lets the attacker read the structure of the victim's access pattern rather than merely detecting its presence.

\parhead{Attacks.} The footprint sensitivity turns the channel into a discriminative attack surface. We build three attacks in two classes, summarized in Figure~\ref{fig:channel-attacks}: a covert channel that requires a colluding sender, and two side-channel attacks against a non-cooperating victim. The covert channel uses calibrated high and low read times to encode bits. The two side-channel attacks measure how long the attacker's read time stays above its idle baseline during generation, then use this duration to infer output length or refusal status.

\parhead{Covert-Channel Attack.}
Unlike the two side-channel attacks, this attack requires a cooperating sender on the victim GPU. The sender encodes each bit by on-off keying its device-to-host DMA: to transmit a~1, it issues a large-footprint burst that spans many distinct pages for one symbol period, and to transmit a~0, it idles. The receiver runs the self-timed probe and thresholds the median latency within each symbol window. A short preamble of known bits calibrates the on-off polarity, after which subsequent symbols are decoded by comparing each window's median against the learned threshold. The footprint sensitivity is what makes the encoding reliable: a burst that touches many pages drives a strong latency shift, whereas idling or reusing a small region does not. A colluding pair can thereby exfiltrate a key or short secret across the GPU boundary at tens of bits per second, crossing process, container, and user isolation (\S\ref{sec:covert}).

\parhead{Length-Inference Attack.}
This attack targets a non-cooperating victim. A passive co-tenant on a second GPU that shares the memory controller observes the duration of generation-phase contention from a co-resident language model whose CPU-offloaded deployment produces sustained device-to-host DMA traffic. The attacker sends the victim nothing and has no access to its API. When the model generates tokens, device-to-host transfers of activations and key-value data contend with the probe and lift its latency above the idle baseline. When generation ends, the latency returns to baseline. To identify individual requests, the attacker segments the continuous latency trace with a threshold detector that computes a local baseline from its own trailing probe history: a sustained excursion marks the onset of a generation, and the return marks its end. Each detected interval yields a measured duration. Because a longer output requires more decoding steps, the generation-phase duration grows with token count, and the attacker calibrates this monotonic relationship with a duration-to-length model trained on one victim process, then applies it blind to a fresh process. Segmentation is the fragile step: baseline drift or closely spaced requests can cause under-segmentation or over-segmentation, accounting for the roughly one-third of fresh victim processes on which end-to-end inference fails. Output length reveals the structure of the victim's response and can distinguish short answers from long explanations or enable downstream attacks that correlate length with known prompt templates (\S\ref{sec:profiling}).

\parhead{Refusal-Inference Attack.}
This attack uses the same duration signal and segmentation procedure as the length-inference attack but makes a coarser binary decision: whether the model refused a query. It requires an additional behavioral assumption, namely that the victim model produces systematically shorter outputs for refusals than for accepted responses. This holds when the model emits a brief policy-violation message against a multi-paragraph answer, but fails on models that answer policy-violating prompts at full length. Given this assumption, the attacker classifies each detected generation interval as ``refused'' or ``accepted'' by thresholding its duration, with the threshold set from a calibration run or from the bimodal structure of the observed durations. Refusal status is a semantic property of the victim's private input: where the assumption holds, the attacker learns that a co-tenant submitted a prompt that the model treated as policy-violating, without any access to the model's API or outputs (\S\ref{sec:profiling}).

Controlled sustained-DMA workloads separately characterize duration and coarse-footprint recovery (\S\ref{sec:robust}).
The underlying channel crosses the process, container, user, and CPU-to-GPU boundaries, since these namespaces do not isolate the shared host memory subsystem. We demonstrate each attack in the cross-GPU process setting and do not re-run every attack across every boundary.

\section{Experiments}
\label{sec:eval}
\subsection{Experimental Setup}
\parhead{Platforms.} Our mechanism and channel-characterization experiments run on the Workstation, which holds two NVIDIA RTX~6000 Ada GPUs on separate PCIe root ports, one Intel Xeon w7-3565X socket, no NVLink, no MIG, and no vGPU, with the IOMMU in translating mode. The real-victim attacks and the cross-NUMA test run on the 2-socket server, which carries four NVIDIA RTX~PRO~5000 and two RTX~PRO~6000 Blackwell GPUs across two sockets. The production stack and the earlier cross-NUMA measurement run on the Rental, a four-GPU server with NVIDIA RTX~PRO~6000 Blackwell GPUs. Each result names its platform.

\parhead{Protocol and Metrics.} The attacker is unprivileged, and hardware performance counters are disabled by the default admin-only policy. Condition order is randomized within each run, and every run was verified free of unrelated GPU jobs. Because the absolute probe latency depends on the load of the attacker, we compare conditions only within a run. Every AUC is a rank-based ROC AUC of the conditioned probe against the idle probe, and unless stated otherwise, it measures small-footprint against large-footprint discrimination. Medians are taken over at least eight trials. We read significance as whether the 95\% bootstrap interval of an AUC excludes 0.5, and we resample independent pairs or windows rather than correlated within-trial samples. We do not place an interval on the folded statistic $\max(\mathrm{AUC},1-\mathrm{AUC})$, whose interval cannot cross 0.5. For paired back-to-back measurements that drift, we additionally report the drift-robust median shift in nanoseconds, and we distinguish the fraction of positive pairs, which is an effect size, from the ROC AUC. Raw per-trial data and the analysis scripts for all experiments are archived.

We organize the evaluation around six questions.
\begin{itemize}
\item[\textbf{RQ1}] Does the cross-GPU channel exist, and what mechanism drives it? (\S\ref{sec:mech})
\item[\textbf{RQ2}] What isolation boundaries does it cross? (\S\ref{sec:reach})
\item[\textbf{RQ3}] Can a colluding pair communicate over the channel? (\S\ref{sec:covert})
\item[\textbf{RQ4}] What can a passive attacker learn about a co-resident language model? (\S\ref{sec:profiling})
\item[\textbf{RQ5}] Does the channel reach production serving stacks? (\S\ref{sec:prod})
\item[\textbf{RQ6}] How robust is the channel, and how much information does it carry? (\S\ref{sec:robust})
\end{itemize}

\subsection{Channel Existence and Mechanism (RQ1)}
\label{sec:mech}
We establish the channel's existence, characterize its sensitivity to the victim's memory footprint rather than bandwidth, and narrow the mechanism through a series of controls, including an \texttt{iommu=off} on/off/on cycle and a cross-agent localization test on the Workstation.

\begin{figure*}[t]\centering
\includegraphics[width=\linewidth]{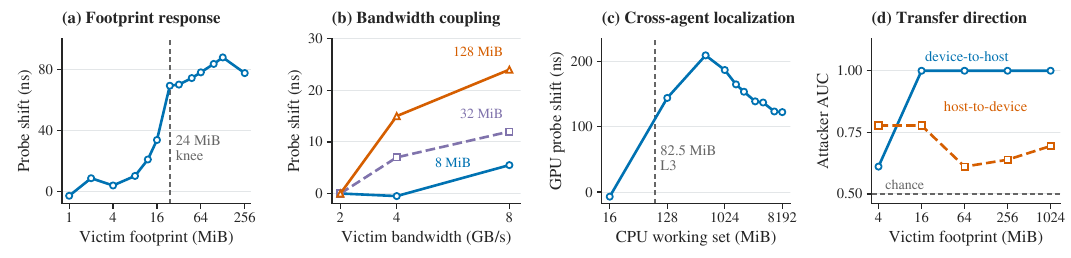}
\caption{Channel mechanism. (a) Footprint dose-response. (b) Bandwidth and footprint, with slopes of 1.0, 1.9, and 3.8~ns per GB/s for 8, 32, and 128~MiB, respectively. (c) Cross-agent localization. (d) Direction asymmetry.}\label{fig:mech}
\end{figure*}

\parhead{Existence and Structure.} At a fixed bandwidth of 4~GB/s, a large-footprint victim transfer and a small-footprint one separate at an AUC of 1.000 with a median shift of 80~ns. This core discrimination holds at an AUC of 1.000$\pm$0.000 across five independent runs. A small-footprint transfer at 4~GB/s against an idle victim scores a raw AUC of 0.29, so a direction-reversing attacker obtains an advantage of $\max(\mathrm{AUC},1-\mathrm{AUC})=0.71$. The small footprint thus carries at most a weak signal of the opposite sign, with the probe running slightly faster, and it does not reach the strong same-direction regime of the large footprint. We have no independent replication of this weak reversed effect and do not build on it. Three negative controls, an SM-bound compute kernel, an allocator and driver storm, and the small-footprint transfer, all score at an AUC of at most 0.55. On this platform, these controls argue against compute contention, CPU-side driver serialization, and bandwidth alone as the cause of the separation.

\parhead{Empirical Model.} Sweeping the footprint at a fixed bandwidth of 4~GB/s produces a monotonic and saturating dose-response with a knee near 24~MiB, as Figure~\ref{fig:mech}(a) shows: the signal grows with the number of distinct host-memory pages the victim touches, not merely with whether a transfer is present. A grid over bandwidth and footprint in Figure~\ref{fig:mech}(b) shows that the two parameters interact: at the smallest footprints, no tested bandwidth produces a measurable signal, so bandwidth alone is insufficient to open the channel. As the footprint grows, bandwidth begins to modulate the signal, with the bandwidth-sensitivity slope rising from 0.9~ns per GB/s at 8~MiB to 3.8~ns per GB/s at 128~MiB. Footprint thus gates the channel and provides discriminatory information that bandwidth alone does not supply. These observations are consistent with the empirical model that relates the signal to the two transfer parameters,
\begin{equation}
\text{signal} \approx k \cdot \text{bandwidth} \cdot f(\text{footprint}),
\end{equation}
in which the footprint term rises from near zero for a single reused row toward a saturating value past the knee. We interpret $f$ as the density of page-table walks that the transfer induces, which rises as the footprint spans more distinct pages and agrees with the \texttt{iommu=off} result below. The fit alone does not establish a mechanism, and the controls that follow narrow it to the IOMMU page-walk path. The separations reported above, including the AUC of 1.000 in the existence test, are measured under controlled conditions in which the attacker knows the victim's active and idle windows and the victim runs a synthetic DMA kernel with a fixed footprint. Under an end-to-end attack on a real LLM workload, where the attacker lacks this knowledge and the victim's DMA pattern is irregular, the achievable accuracy is substantially lower. Section~\ref{sec:llm} reports these label-free numbers.

\parhead{Localization to Host DRAM.} A CPU memory stressor, rather than a GPU, drives the same signal into the GPU probe, and the signal activates when the working set of the stressor exceeds the CPU last-level cache of 82.5~MB, which is the point at which the stressor begins to hit DRAM. Figure~\ref{fig:mech}(c) shows this behavior, with a latency shift of up to 210~ns. On this platform, the result excludes any GPU-internal structure and localizes the contended resource to the shared host memory subsystem. It also motivates the prediction, which we leave to future work, that any DMA agent that shares the memory controller could drive or observe the signal just as the GPUs do.

\parhead{IOMMU Control.} The signal is direction specific and configuration specific. Only pinned-memory DMA produces a signal, while pageable and bounce-buffered DMA does not (AUC~0.41). Sequential and random access at an equal footprint behave similarly, and small copies over a large footprint already produce a signal, so the effect does not require large individual transfers. A page-size lever that compares 4~KiB pages against 2~MiB transparent huge pages leaves the plateau height unchanged. This result is inconclusive, because we could not verify that the driver programmed 2~MiB IOMMU mappings, and enabling transparent huge pages on the CPU side need not change the IOMMU page size or the number of distinct IOMMU translations. The strongest control is \texttt{iommu=off}, which we run as a replicated on/off/on cycle. With the IOMMU translating, the large-footprint separation from an idle victim reaches an AUC of 0.87 to 1.00 with a shift of 70 to 79~ns across two on-state runs. With the IOMMU disabled, the separation collapses to chance (AUC 0.57, shift -0.5~ns), and when the IOMMU is restored, it returns (AUC 1.00, shift 70~ns), over fifteen trials per condition in each state. The small-footprint condition never separates in any state. The idle baseline drops by about 116~ns when translation is disabled and returns when it is restored, which independently confirms that the toggle took effect, and the on/off/on reversal rules out drift or a one-off event. A matched-bandwidth distinct-page contrast points to a page-walk origin. At a fixed 4~GB/s, a small footprint that touches few distinct pages never leaks, while a large footprint that touches many pages leaks only when the IOMMU is translating, so the channel scales with the number of page-table walks that the transfer induces rather than with bandwidth. Because the two GPUs sit behind separate DMAR units with separate IOTLBs, a shared translation cache is excluded by topology. The channel is therefore IOMMU-dependent and consistent with page-table-walk descriptor contention in shared host DRAM rather than with DRAM row buffers. The control does not by itself exclude every change in the DMA data path that disabling the IOMMU also induces, such as a change in physical addressing, and an \texttt{iommu=pt} passthrough control or a control that holds total DMA bytes fixed while varying the number of distinct pages would separate them further. A direct DRAM bank-mapping attempt separately did not resolve a clean bank structure on this host.

\parhead{Direction and the Real Workload.} A matched-footprint sweep resolves the direction of the channel. We hold bandwidth fixed and sweep the footprint from 4 to 1024~MiB in each direction, and device-to-host writes show a footprint dose-response in Figure~\ref{fig:mech}(d). Their AUC climbs from 0.65 at 4~MiB to 1.00 from 16~MiB upward, with a median shift that grows to 83~ns, while host-to-device reads stay within 3~ns of the baseline at every footprint and show no footprint dependence. Host-to-device reads still reach an AUC near 0.79 at the smallest footprints on few samples, so we describe them as lacking the strong footprint-dependent regime rather than as strictly non-leaking. At a matched 256~MiB footprint, the device-to-host shift is 80~ns against a host-to-device shift below 1~ns, and two independent clean runs agree. The channel is therefore a device-to-host write channel. This result resolves an apparent tension with the real workload. We instrument the PCIe traffic of the victim during generation and measure about 8~GB/s of host-to-device weight streaming against about 1~GB/s of device-to-host transfer of activations and key-value data, a ratio near eight to one. The victim is thus dominated by host-to-device traffic, yet it leaks, and its leak is weak, near 6~ns (Section~\ref{sec:llm}). Because host-to-device transfers do not couple to the channel at any footprint, the signal is carried by the roughly 1~GB/s device-to-host component. The probe does not track the instantaneous transfer rate, since it responds to the presence of device-to-host activity and the shift is small against its own noise, so the attribution combines the measured device-to-host volume with the controlled direction sweep. A victim whose device-to-host and host-to-device volumes vary independently would isolate the direction on the victim itself, which remains future work.

\begin{answerBox}
\textbf{Answer to RQ1.} Under controlled conditions the channel exists (AUC 1.000 across five runs) and is footprint-sensitive: the signal scales with the number of distinct pages the victim touches. Footprint gates the channel, and bandwidth alone is insufficient to open it. A replicated \texttt{iommu=off} control traces the mechanism to IOMMU page-walk contention in shared host DRAM, and the channel is a device-to-host write channel. End-to-end attack accuracy on real workloads is lower (Section~\ref{sec:llm}).
\end{answerBox}

\begin{table}[t]\centering\footnotesize
\caption{Isolation boundaries and mechanism controls.}\label{tab:boundaries}
\setlength{\tabcolsep}{4pt}
\begin{tabular}{@{}lll@{}}
\toprule
Boundary or control & Crossed & AUC (evidence) \\
\midrule
Separate process & yes & 1.000 \\
Docker container & yes & 1.000 \\
Separate GPU without NVLink & yes & 1.000 (5 runs) \\
Unprivileged Unix user & yes & 0.955 (2 windows) \\
Separate memory controller & no & attenuated \\
IOMMU disabled & no & 0.57 off, 1.00 on \\
\bottomrule
\end{tabular}
\end{table}

\subsection{Cross-Boundary Reach (RQ2)}
\label{sec:reach}
Having established the channel and its mechanism, we test which isolation boundaries it crosses and where it attenuates. We probe it across process, container, and Unix-user boundaries, and measure its locality with a cross-NUMA placement that pins both parties to known controllers.

\parhead{Isolation Boundaries.} Table~\ref{tab:boundaries} summarizes the boundaries that the channel crosses. In this subsection, the AUC measures the separability of a probe reading taken while the victim is active from one taken while it is idle, under conditions that we control. It is the figure that a covert receiver or a defender would use, but it presumes knowledge of the victim's active and idle windows, so it bounds what a passive attacker can achieve from above, and Section~\ref{sec:llm} reports the achievable, label-free number through a causally held-out online detector. The channel scores 1.00 from the first GPU to the second, 0.88 in the reverse direction, and 1.000 for an attacker inside a Docker container. A CPU memory stressor drives the GPU probe in the CPU-to-GPU direction, as Figure~\ref{fig:mech}(c) shows, and a GPU sender drives a CPU probe at an AUC of 0.93, so container namespaces do not isolate the shared host memory subsystem. The channel also crosses Unix-user isolation on the Rental platform. With the attacker probe running as a separate unprivileged user, holding no root privilege, no performance counters, and no shared context with the root victim, device-to-host writes are detected at an AUC of 0.955. This value matches the 0.960 of a same-user control, so any co-tenant user can mount the attack. Two windows support this result, since an earlier window on a different probe GPU detects the unprivileged attacker at an AUC of 0.867 against a same-user control of 0.910, and both windows separate the victim from idle.

\parhead{Memory-Controller Locality.} A cross-NUMA measurement pins both the probe and the victim to a memory controller, so that their host buffers land on a known node, and it shows that the channel is local to the controller. The decisive test is a device-order-verified run on the 2-socket server, with each GPU confirmed by its PCIe address and NUMA node. A same-controller victim shifts the probe median by 336~ns above an 845~ns idle baseline, while a victim on the other controller shifts it by 0.3~ns, over forty pairs per condition with no overlap between the two conditions. A symmetric control that places the probe on the second socket gives the same pattern, a 381~ns same-controller shift against 0.3~ns across controllers. This control also rules out a silent victim as the cause of the cross-controller null, since the same victim GPU produces a large shift when it is probed from its own controller. An earlier measurement on the Rental platform, which was not device-order-verified, agrees qualitatively, with a same-controller AUC of 0.955 (95\% interval [0.90, 1.00], excluding 0.5) against a cross-controller AUC of 0.551 (interval [0.40, 0.70], including 0.5) over twenty-four pairs. Co-location on the same memory controller is therefore necessary for the channel, which supports a shared-memory-controller mechanism and a placement mitigation that puts mutually distrusting tenants on separate memory controllers.

\begin{answerBox}
\textbf{Answer to RQ2.} The channel crosses process, container, and Unix-user boundaries (AUC 0.955 to 1.000). It is local to the memory controller: a same-controller victim shifts the probe by hundreds of nanoseconds, while a cross-controller victim shifts it by a fraction of a nanosecond.
\end{answerBox}

\subsection{Covert-Channel Attack (RQ3)}
\label{sec:covert}
We test whether a colluding pair can exploit the channel for deliberate communication across the GPU boundary.

A colluding sender on one GPU transmits to an unprivileged receiver on another by on-off keying its device-to-host bursts, one symbol per window, and the receiver decodes the median probe latency per window. On the 2-socket server, the channel is reliable at tens of bits per second. A random 512-bit stream at a 50~ms symbol period, which is 20 bits per second, decoded with zero bit errors, and the rule of three gives a 95\% upper bound of about 0.6\% on the bit-error rate. A device-order-verified replication confirms this result and maps the reliable raw rate. Three fresh 200-bit transmissions at 20 bits per second decode at a raw bit-error rate of 1\% to 2\%, and one at 50 bits per second decodes at 0.5\%, while at 100 bits per second the raw rate rises to about 9\%, which is usable only with an error-correcting code. The on-off polarity, which defines the latency level that denotes a one, is calibrated from a short training preamble. Before that calibration, the first cold-start transmission decoded at chance, about 0.48, which we disclose as a required warm-up instead of folding it into the rate. The probe sample rate under contention sets the limit. The bursts of the sender slow the self-timed probe from about 800 to about 340 samples per second, which leaves roughly seventeen samples per bit at 20 bits per second but only about three at 100. The demonstrated reliable raw rate is therefore tens of bits per second, which suffices to exfiltrate a key or a short secret.

\begin{answerBox}
\textbf{Answer to RQ3.} Yes. A colluding pair communicates at 20 to 50 bits per second with a raw bit-error rate of 0 to 2\%, sufficient to exfiltrate a key or short secret.
\end{answerBox}

\subsection{Generation Profiling Attacks (RQ4)}
\label{sec:profiling}\label{sec:llm}
We turn the channel against a real co-resident language model. A passive attacker who sends the victim nothing uses the probe to detect generation activity, recover the per-request output length, and infer whether the model refused a query, which is a semantic property of the victim's private input.

\begin{figure}[t]\centering
\includegraphics[width=\columnwidth]{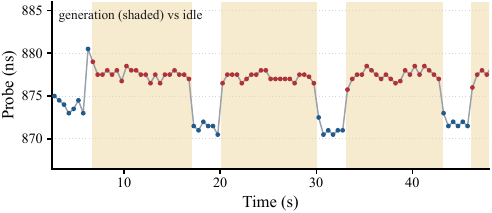}
\caption{Activity detection on a real CPU-offloaded language model. The 0.5~s probe medians track the shaded generation windows against idle intervals, at a per-bin AUC of 0.888.}
\label{fig:phase}
\end{figure}

\parhead{Activity Detection.} On a real CPU-offloaded language model, the probe distinguishes generation from idle intervals, as Figure~\ref{fig:phase} shows. Over twenty generation requests and 532 half-second bins, the threshold-free per-bin AUC is 0.888 against a 76.3\% always-active baseline. This AUC is our primary measure, because it does not depend on a threshold. For an online decision, we also report a temporally held-out threshold, fixed on the first ten requests and tested on the last ten. It uses a local estimator that subtracts a trailing-only rolling baseline, which the attacker computes from its own past probe stream, so the detector uses no future samples and is realizable online. This temporally held-out test reaches a balanced accuracy of 0.885 and detects nine of the ten held-out requests, at a held-out AUC of 0.824. An interleaved split that shares drift is more optimistic at 0.956, which brackets the gap that the temporal split closes, and we release the analysis script that fixes these choices. The generation-window latency shift is small in absolute terms, about 6~ns on an 872~ns baseline, so the per-sample signal is modest, and half-second binning is what lifts detection, with the generation request as the primary independent unit. The probe baseline drifts over the several-minute trace, so a global threshold underperforms, while the AUC and the local drift-corrected estimator of Section~\ref{sec:robust} are unaffected. On this workstation trace, the per-sample shift is too small to resolve token counts. On the higher-coupling platform that we describe next, the same probe recovers the per-request output length and a semantic secret of the private input of the victim.

\begin{table*}[t]\centering\footnotesize
\caption{Real and production victim results ($\rho$, held-out Spearman correlation under clean segmentation).}\label{tab:victims}
\setlength{\tabcolsep}{3pt}
\begin{tabular}{@{}llllll@{}}
\toprule
Task & Runtime & Model & Device-to-host regime & Result & Verdict \\
\midrule
Output length & transformers & 3 families, 0.5--14B & offloaded, loaded & $\rho$ 0.95--0.998, end-to-end 6/9 & leaks \\
Output length & SGLang & Qwen2.5-14B & weight offload, 6~GB/s & $\rho$ 0.996 & leaks \\
Refusal & transformers & Qwen2.5-14B & offloaded, loaded & balanced accuracy 0.65--0.82 (2/3) & leaks \\
Refusal & transformers & Qwen2.5-0.5B & offloaded, loaded & balanced accuracy 0.955 & leaks \\
Refusal & transformers & Llama-3.1-8B & offloaded, loaded & over-segmentation or chance & unreliable \\
Refusal & transformers & Mistral-7B & refuses 2/22, gap $\sim$0 & no duration signal & no leak \\
Covert channel & none & none & on-off keyed & bit-error rate 0--2\% at 20--50 bit/s & reliable \\
Activity & transformers & 0.5B & 500~MB/s sustained & per-bin AUC 0.73 & weak leak \\
Activity & transformers & 1.5B & 800--960~MB/s sustained & per-bin AUC 0.888, online 0.885 & leaks \\
Activity & transformers & 7B & slow, $\sim$1.3~s per token & per-bin AUC 0.53 & no leak \\
Activity, cache hit & vLLM+LMCache & 7B & bursty, median $\sim$0 & AUC 0.497 & no leak \\
Activity, all miss & vLLM+LMCache & 7B & bursty & AUC 0.495 & no leak \\
Activity, batch 48 & vLLM+LMCache & 7B & 8.47~GB/s peak, bursty & AUC 0.495 & no leak \\
Activity, 256~MiB pool & vLLM+LMCache & 7B & 1.0~GB/s sustained & AUC 0.496 & no leak \\
Cache hit or miss & vLLM+LMCache & 7B & per request & held-out AUC 0.522 & no secret \\
\bottomrule
\end{tabular}
\end{table*}

\parhead{Output-Length Inference.} On the 2-socket server, the channel yields a meaningful secret. We run a CPU-offloaded language-model victim on one GPU and an unprivileged self-timed probe on another, train a duration-to-length model on one victim process, and test it blind on a fresh victim process. In these experiments, the output length of each request is experimentally controlled to a known value, and this controlled setting is our primary and cleanest measurement. The attacker recovers the per-request output length from the duration signal at a held-out Spearman correlation of 0.99 on a fresh process when its requests segment cleanly. Conditional on clean segmentation, this recovery generalizes across model families and scales, reaching 0.95 to 0.998 on Qwen2.5 at 0.5, 7, and 14 billion parameters, 0.997 on Mistral-7B-v0.3, and 0.959 on Llama-3.1-8B, so it is a property of the channel rather than of one model. We also test a natural-generation variant in which the model terminates on its own end-of-sequence token (SGLang chat completions, a 256-token cap, and no minimum). Over a workload that spans 7 to 256 tokens, of which about a third terminate naturally and the rest reach the cap, the channel still recovers output length on a fresh process at a held-out Spearman correlation of 0.71 to 0.80. This correlation is weaker than the near-0.99 of controlled lengths, and like it, the variant ties the utilization baseline on the three-class task. An earlier raw-completion attempt, dominated by cap-truncated outputs, is a negative result in the artifact.

\begin{figure*}[t]\centering
\includegraphics[width=0.96\linewidth]{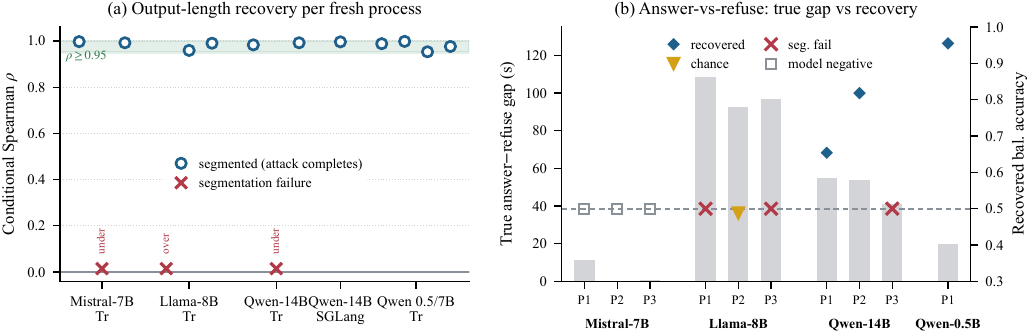}
\caption{Language-model attacks over all fresh-process attempts. (a) Output-length recovery. (b) Answer-versus-refusal outcome.}
\label{fig:llm}
\end{figure*}

\parhead{End-to-End Reliability.} We separate this conditional fidelity from end-to-end reliability. We score every predeclared fresh test process with the frozen detector and count as an attack failure any process whose detector output is not a sane segmentation, either because fewer than half the requests match (under-segmentation) or because it detects more than twice as many runs as there are requests (over-segmentation). Under this criterion, the attack completes on about two thirds of fresh processes, six of nine across the three larger models, each at $N{=}3$. The failures are of both kinds, namely under-segmentation when the probe baseline drifts during a long process and gross over-segmentation that fragments generations and destroys the duration-to-length map. The channel thus carries output length at high fidelity when segmentation succeeds, but its per-process reliability against a blind victim is imperfect, and we report every process instead of only the successful ones, as Figure~\ref{fig:llm} shows over all fresh-process attempts. On the three-class task of short, medium, and long outputs, the duration model ties a coarse utilization baseline on most models and beats it only on Llama, so it fails our pre-registered composite criterion for beating utilization. Our claim is therefore that the timing channel carries output length, and we make no claim that it beats a simpler activity proxy.

\parhead{Refusal Inference.} From the same duration signal, the attacker recovers whether the model refused a query when the model emits a short refusal against a longer answer. This inference is both model-dependent and process-variable. It is recoverable end-to-end on Qwen2.5-14B on two of three fresh processes (balanced accuracy 0.65 and 0.82, with the third over-segmenting) and on a small Qwen2.5-0.5B process (0.955 over forty-four balanced prompts). It is unreliable on Llama-3.1-8B. Although the true generation-time gap between answer and refusal there is large, about ninety to one hundred twenty seconds, one of three fresh processes over-segments, one under-segments, and the single one that segments cleanly lands at chance. It is absent on Mistral-7B, whose answer and refusal generation times differ by about zero seconds on all three fresh processes, because the model answers policy-violating prompts at full length instead of emitting short refusals. The attack recovers refusal only when the victim both produces systematically shorter refusals and segments cleanly. Refusal here is a property of the behavior of the victim model on the evaluated prompt set, so where it holds, the attacker learns that a co-tenant submitted a prompt that the model treated as policy-violating. Both inferences need the shared host memory subsystem to be under load, which amplifies the device-to-host traffic of the victim into a detectable few-nanosecond probe shift. We do not assume that realistic co-tenancy always supplies this load. The attacker can generate the load at an observable resource cost, and on an otherwise-idle host, it lifted detection from zero to all thirty windows and output-length recovery to a correlation of 0.996.

\parhead{A Second Serving Runtime.} The positive results above use a \texttt{transformers} device-map CPU-offload victim. To test whether the leak is a property of that runtime or of the host-transfer regime that it produces, we repeat the output-length attack against a second, independently implemented runtime, SGLang with its own native CPU weight-offload path (\texttt{--cpu-offload-gb}). We first characterize its regime with a PCIe throughput counter on the victim GPU (\texttt{nvidia-smi dmon -s t}, where Rx is host-to-device and Tx is device-to-host). During generation, it sustains about 58~GB/s of host-to-device weight streaming and about 6~GB/s of device-to-host traffic at a near-unit duty cycle, which places it inside the device-to-host regime that the channel exploits. We report sustained bandwidth and duty cycle for this victim because its transfer footprint is not directly observable. The frozen attack places the probe on a second GPU on the same memory controller, with the requests of the victim separated by idle gaps long enough for the frozen detector to segment them. It recovers output length at a held-out Spearman correlation of 0.996 with full coverage on a fresh process, and it ties the utilization baseline on the three-class task as before. The attack therefore transfers across two serving runtimes under sustained device-to-host traffic, and the host-transfer regime that the deployment produces governs exploitability more than the runtime above the channel does. Two runtimes support full runtime independence without proving it. A request-spacing sweep bounds the segmentation side of the attack. Recovery is clean when requests are at least a couple of seconds apart, but at inter-request gaps of one second or less, the generations merge under the detector and per-request recovery fails. With two concurrent generations, the detector resolves only two runs across sixty requests. The demonstrated attack therefore applies to sequential, gap-separated requests, and concurrent or high-rate serving would need a receiver that models overlapping decodes, which we leave to future work.

\parhead{Workload Scope.} We repeat the real-victim measurement across offload configurations. Fast-generating models leak, and among them, detection strengthens with device-to-host bandwidth, from an AUC of 0.73 for a 0.5-billion-parameter model at 500~MB/s to an AUC of 0.94 to 0.98 for a 1.5-billion-parameter model at 800 to 960~MB/s. A 7-billion-parameter model heavily offloaded to the CPU does not leak, at an AUC of 0.53, despite a comparable average device-to-host rate. It generates slowly at about 1.3~s per token, so its DMA is spread thin, and its multi-second generations expose the detector to baseline drift. The channel therefore favors victims with concentrated device-to-host activity, while slow, heavily offloaded large models make poor victims. Table~\ref{tab:victims} collects these victims, where every figure is held out on a fresh process and conditional on clean segmentation of the requests.

\begin{answerBox}
\textbf{Answer to RQ4.} A passive attacker detects generation at a per-bin AUC of 0.888 and recovers per-request output length at a Spearman correlation of 0.99 across three model families, conditional on successful request segmentation. End-to-end segmentation succeeds on about two thirds of fresh processes. The remainder fail at baseline drift or closely spaced requests. Refusal inference reaches a balanced accuracy of 0.955 but is model-dependent, since it requires a victim model that produces systematically shorter refusals, which not all models do.
\end{answerBox}

\subsection{Production Stack and Leakage Regime (RQ5)}
\label{sec:prod}
The attacks above use CPU-offloaded models with sustained device-to-host traffic. We now test whether the channel reaches a production serving stack whose traffic pattern differs, deploying an unmodified vLLM+LMCache key-value offload under escalating regimes and mapping the leakage boundary with a synthetic phase diagram.

\parhead{Production Stack.} We test an unmodified production system instead of an offload proxy, and we deliberately try to elicit a leak. A real vLLM 0.9.2 server with LMCache~0.5.4~\cite{lmcache} offloads its key-value cache to host memory while it serves a 7-billion-parameter model on the Rental platform. It does not leak across four escalating regimes, and every run was verified free of co-tenant activity before and after. Cache-hit-dominated and write-heavy all-miss serving score threshold-free AUCs of 0.497 and 0.495. Raising concurrency to a continuous 48-request batch lifts the measured device-to-host peak to 8.47~GB/s, into the range where our synthetic trigger leaks strongly, yet detection stays at 0.495. Shrinking the offload pool to 256~MiB forces the writes into the small-footprint regime that leaks on this platform, at a mean device-to-host rate of 1.0~GB/s with sustained eviction, and it still scores 0.496. Peak device-to-host bandwidth alone therefore does not explain the negative. A candidate explanation is that production key-value offload writes in short chunks between compute steps, about 16 milliseconds at a time, so its device-to-host traffic is bursty with a median rate near zero. The SGLang weight-offload victim, whose offload is sustained, does leak, but it ran on the 2-socket platform, so the contrast is cross-platform. The two results are consistent with exploitability that tracks the host-transfer regime, but we do not treat them as a same-hardware proof that burstiness alone explains this negative.

\parhead{Burstiness and Average Pressure.} A controlled synthetic experiment on a single platform supports the regime reading without fully isolating it. A continuous 50~GB/s peak device-to-host stream discriminates from idle at an AUC of 0.85, while the same peak delivered in 16-millisecond bursts at a 16\% duty cycle stays near chance at 0.54. These two conditions differ in both persistence and time-averaged bandwidth, about 50 against 8~GB/s, so they do not separate intermittency from average memory pressure. A matched-mean comparison of sustained against bursty traffic at the same 8~GB/s average points the same way but weakly, 0.57 against 0.54, and it rests on ten independent windows per condition from a single run. We therefore read the production negative as consistent with low time-averaged host-memory occupancy, whether from burstiness or from low average pressure. Either way, a production single-node key-value offload is a poor victim, which places the channel's operating regime at sustained device-to-host traffic, such as large-scale collective or tensor-parallel spill, and at slower-memory hosts where a moderate victim already leaks.

\parhead{Multi-Tenant Aggregation.} We run one, two, and four independent offload instances with staggered Poisson arrival. The aggregate device-to-host duty cycle rises from 6\% to 68\% as their bursts overlap, yet detection stays at chance, because the aggregate rate reaches only about 0.85~GB/s at four instances, which is below the leakage floor that we measure next. Overlap removes the burstiness but not the low absolute bandwidth, so a single-node multi-tenant offload does not leak either.

\parhead{Leakage Phase Diagram.} A synthetic sweep over mean bandwidth and duty cycle at a fixed footprint, on the more sensitive probe placement, sketches a leakage phase diagram. Because the baseline of the shared instance drifts across the sweep, we score each cell with the drift-robust within-pair median shift between the conditioned and idle probe, in place of a pooled AUC that drift distorts, and we report the fraction of positive pairs alongside as an effect size. No cell leaks below a mean of about 2~GB/s, where the lowest cell has a median shift of +0.6~ns with a 95\% interval that includes zero. Above that floor, the median shift rises with both the mean and the duty cycle. At a mean of 8~GB/s, it grows from +2.2 to +3.8 to +4.5~ns at duty cycles of 25\%, 50\%, and 100\%, with the positive-pair fraction rising from 0.81 to 0.94, and at a mean of 4~GB/s, it rises from +0.6~ns at a 10\% duty cycle to +2.3~ns at a 50\% duty cycle. Leakage therefore strengthens as the device-to-host stream becomes more sustained, which supports reading the production negative, whose traffic is bursty and low in time-averaged rate, as a low-occupancy regime below this floor. Every cell rests on fifteen or sixteen paired trials from a single clean run, several cells show a baseline drift of tens of nanoseconds within the run, and we therefore report the sweep as indicative instead of as a calibrated boundary.

\begin{answerBox}
\textbf{Answer to RQ5.} No. A production vLLM+LMCache key-value offload does not leak across four escalating regimes (AUC 0.495 to 0.497). The channel's operating regime requires sustained device-to-host traffic above about 2~GB/s mean bandwidth. Bursty or low-duty-cycle offload stays below this floor and does not leak.
\end{answerBox}

\subsection{Robustness and Information Content (RQ6)}
\label{sec:robust}
We quantify the channel's information capacity and its robustness to noise, baseline drift, and probe-configuration variation using controlled synthetic workloads on the Workstation.

\parhead{Duration Inference.} On a synthetic sustained-DMA workload with known large and small conditions, the recovered activity duration matches the ground truth with a correlation of 1.0 and a class accuracy of 100\% over twelve conditions, at a mean absolute error of 0.51~s. On the Workstation, token-count inference is robust only under this sustained-DMA regime, and on intermittent weight-streaming offload, the same estimator is unreliable, with a correlation between 0.28 and 0.49. This workstation limit does not carry to the 2-socket server, where output-length inference on a real offloaded victim reaches a correlation of 0.99 (Section~\ref{sec:llm}).

\parhead{Information Content.} The single-observation mutual information between footprint and the probe is 0.86 bits of a possible 2.32, and a five-class footprint classifier reaches 42\%, which is 2.1$\times$ the 20\% baseline. The three largest footprints are mutually confused, because the signal saturates past the knee, so the channel behaves as a low-cardinality discriminator of the victim footprint.

\parhead{Robustness.} Discrimination holds at an AUC of 1.0 under heavy background CPU-DRAM noise and stays between 0.83 and 1.0 across probe buffer, stride, and step configurations. The absolute probe baseline drifts by 250 to 400~ns over a minute, which exceeds the signal, so a naive across-time measurement degrades to an AUC of 0.67. An interleaved and paired estimator, which measures the two conditions back to back and compares them, cancels the drift and recovers 100\% in this offline setting. This paired procedure requires that both conditions be observed, so it supports the offline covert and controlled-victim results and does not by itself establish an online attack that uses only attacker-controlled operations. Section~\ref{sec:limits} summarizes how much replication backs each of these results.

\begin{answerBox}
\textbf{Answer to RQ6.} Under controlled conditions, the channel recovers activity duration at a correlation of 1.0 and carries 0.86 bits per observation. Discrimination holds at AUC 1.0 under heavy CPU-DRAM noise. Baseline drift degrades naive measurement (AUC 0.67), but an interleaved paired estimator cancels it.
\end{answerBox}

\section{Defenses}
\label{sec:defense}
The mechanism analysis in Sections~\ref{sec:mech} and~\ref{sec:reach} yields three insights that each suggest a defense.
First, the signal is footprint-sensitive: it tracks the number of distinct host pages the victim touches rather than the bandwidth, so equalizing the footprint across secrets removes the distinguishing signal.
Second, the signal is IOMMU-dependent: it arises from page-walk descriptor reads in shared host DRAM, so reducing the number of distinct translations that a DMA stream induces weakens it at the source.
Third, the signal is memory-controller-local: a same-controller victim shifts the probe by hundreds of nanoseconds while a cross-controller victim shifts it by a fraction of a nanosecond, so placing mutually distrusting tenants on distinct controllers removes the shared substrate altogether.

These observations yield two complementary classes of defense.
Signal-level mitigations leave the shared substrate in place but deny the attacker a discriminative measurement. Footprint flattening and fewer distinct translations fall into this class.
Substrate-level mitigation eliminates the shared resource itself by separating tenants onto distinct memory controllers.
We evaluate the first and third below. The second is a candidate that we have not yet measured.

\parhead{Signal-Level: Footprint Flattening.} In footprint flattening, the victim pads the footprint of the smaller secret up toward the larger one to hold a constant working set. We measure the advantage of the attacker as $\max(\mathrm{AUC}, 1-\mathrm{AUC})$, since an attacker is free to invert its decision, and we add an identical-condition null control, in which the padded small secret equals the large secret, to quantify the finite-sample floor of this estimator. Over about thirty-three trials per condition, the advantage falls from 0.94 without padding to 0.76 at a 64~MiB pad and to 0.53 at a 256~MiB pad, while the null control reads 0.55. By a 256~MiB pad, the advantage has therefore reached the measurement floor, so the secret is no longer distinguishable from an identical workload, and footprint flattening drives the channel into the noise at that overhead. We also evaluate an adaptive attacker that retrains both its threshold and its decision direction on a training split collected under the active defense and is scored on a held-out split. Its balanced accuracy falls from 0.96 without padding to 0.70 at a 64~MiB pad and to 0.47 at a 256~MiB pad, with a 95\% interval of [0.35, 0.56] that brackets the 0.49 null control, so even a retrained attacker is reduced to chance at a 256~MiB pad. The overhead is the cover footprint that the victim must sustain, a constant 256~MiB working set and the memory bandwidth to touch it. A full accounting of the resulting throughput, latency, and energy cost on a production workload remains to be done, so we do not claim that the defense is low-overhead. Constant-footprint cover alone is insufficient, since bandwidth still modulates the signal once the footprint is large.

\parhead{Signal-Level: Fewer Distinct Translations.} Given the IOMMU translation mechanism, a candidate mitigation reduces the number of distinct translations that the DMA of the victim induces. The victim can confine its working set to few pages or back its DMA buffers with huge pages, so that a given footprint spans fewer IOMMU entries. This mitigation targets the footprint term that drives the signal, and as we observe, sequential and random access at an equal footprint behave alike. We have not measured this variant.

\parhead{Substrate-Level: Memory-Controller Placement.} Separating mutually distrusting tenants onto distinct memory controllers nearly eliminates the channel and removes the shared substrate. Its cost is the placement constraint itself and any cross-controller memory-latency penalty that a co-scheduled tenant would otherwise avoid, which we have not quantified. Its correctness rests on the memory controller being the contended resource, which the device-order-verified cross-NUMA test of Section~\ref{sec:reach} confirms in both socket directions, with a same-controller victim shifting the probe by hundreds of nanoseconds and a cross-controller victim shifting it by a fraction of a nanosecond. An operator should still confirm the host topology, since the defense assumes that tenants can be pinned to known controllers.

\section{Related Work}
\parhead{GPU Side Channels.} GPU side channels are practical and have targeted website fingerprinting, user interaction, and neural-network secrets~\cite{naghibijouybari2018rendered}, built covert channels from on-GPU resource contention~\cite{naghibijouybari2017constructing}, recovered cryptographic keys through GPU memory timing~\cite{jiang2016complete}, and stolen a model through context-switch leakage~\cite{wei2020leaky}. The GPU has also served as an engine for microarchitectural attacks on the rest of the machine~\cite{frigo2018grand}, and GPU confidential computing is the emerging hardware defense~\cite{zhu2024confidential,trustmis,ppnas}. These channels operate within one GPU~\cite{invalidatecompare,tunnels,tlbcrossvm}, whereas we report a signal between two GPUs. Cross-GPU channels that require an interconnect use NVLink or DGX peer-to-peer links and cross-GPU L2 contention~\cite{spygpubox,nvbleed}, while our setting has no NVLink and the signal localizes to the host memory subsystem. Host and PCIe bandwidth channels and GPU uncore channels exploit bandwidth and utilization~\cite{lockeddown,invisibleprobe,veiledpathways}, while our signal is footprint-sensitive, and on our platform, bandwidth alone does not drive it. Table~\ref{tab:comparison} summarizes the distinctions along three axes: the shared hardware resource, the attacker's measurement mechanism, and the required hardware assumptions.

\begin{table}[t]\centering\footnotesize
\caption{Comparison with closest prior channels (MC\,=\,memory controller).}\label{tab:comparison}
\setlength{\tabcolsep}{3pt}
\begin{tabular}{@{}lllc@{}}
\toprule
Work & Resource & Probe & Cross-GPU \\
\midrule
DevIOus~\cite{devious} & Shared IOTLB & Eviction & No \\
IOTLB-SC~\cite{iotlbsc} & Shared IOTLB & Set contention & No \\
LockedDown~\cite{lockeddown} & PCIe bus & BW timing & No \\
Inv.\ Probe~\cite{invisibleprobe} & PCIe switch & Congestion & Yes \\
Veiled Path.~\cite{veiledpathways} & GPU uncore & L2/crossbar & No \\
\textbf{Ours} & \textbf{Host MC} & \textbf{Self-timed} & \textbf{Yes} \\
\bottomrule
\end{tabular}
\end{table}

\parhead{Microarchitectural Contention Channels.} Contention channels read a victim through a resource that both parties share. Classic cache attacks use flush-based and contention-based primitives on shared caches~\cite{yarom2014flush,osvik2006cache,gruss2016flush}, reach the last-level cache across cores and virtual machines~\cite{liu2015last,irazoqui2015sa}, and automate leakage profiling~\cite{gruss2015cache}. Beyond caches, the on-chip ring interconnect~\cite{paccagnella2021lord}, execution ports~\cite{aldaya2019port}, processor power and frequency~\cite{wang2022hertzbleed,lipp2021platypus}, and speculative execution~\cite{kocher2019spectre} all carry timing signals, and memory-controller contention is strong enough to serve as a cross-tenant denial-of-service vector~\cite{zhang2017dos}. Our channel is a contention channel in this lineage, and its shared resource is the host memory subsystem beneath two physically separate GPUs instead of a shared CPU structure.

\parhead{DRAM, Memory Bus, and Rowhammer.} DRAM addressing has been reverse-engineered to drive cross-CPU attacks~\cite{drama}, the memory bus carries a high-speed covert channel in the cloud~\cite{wu2012whispers}, and Rowhammer turns DRAM disturbance into a fault primitive~\cite{kim2014flipping,jattke2022blacksmith}, including over the network~\cite{tatar2018throwhammer} and against GPU memory~\cite{gpuhammer}. These studies stay within a single host and a single class of agent or operate on in-GPU GDDR, while our contention arises in shared host DRAM and is driven across the GPU boundary.

\parhead{IOMMU and DMA.} DMA from untrustworthy peripherals bypasses IOMMU protection~\cite{markettos2019thunderclap,morgan2016bypassing}, and the IOMMU is a known source of timing channels. DevIOus~\cite{devious} and IOTLB-SC~\cite{iotlbsc} exploit contention in a shared IOTLB within one IOMMU context. Our two GPUs sit behind separate DMAR units with separate IOTLBs, so a shared translation cache is excluded by topology. A replicated \texttt{iommu=off} control confirms that the signal is IOMMU-dependent yet traces the contention to page-table-walk descriptors in host DRAM rather than to IOTLB eviction (Table~\ref{tab:comparison}).

\parhead{Cloud Co-Residence and Cross-VM Isolation.} Public-cloud tenants can be placed onto the same host~\cite{ristenpart2009hey,varadarajan2015placement}, detected through a side channel~\cite{zhang2011homealone}, and attacked across the virtual-machine boundary to extract keys~\cite{zhang2012cross,cryptgraph}. Our threat model shares this multi-tenant setting, and we show that pinning tenants to separate GPUs does not isolate them.

\parhead{Leakage from ML and LLM Serving.} Machine-learning systems leak through their interfaces, by model extraction~\cite{tramer2016stealing,carlini2024stealing}, membership inference~\cite{shokri2017membership}, and general privacy side channels~\cite{privacysidechannels,splitinference}. Timing leakage now targets language-model serving specifically, recovering response tokens from encrypted traffic~\cite{weiss2024prompt}, stealing inputs through cache and allocation timing~\cite{zheng2024inputsnatch,song2024early}, and reading hardware caches during local inference~\cite{iknowwhatyousaid}. Serving-layer work also targets victim signals from MIG and logical vantage points~\cite{behindbars}. Closest in spirit, SpliceLeak recovers the private input prompt of a victim in a retrieval-augmented generation stack by timing the fusion boundary of non-prefix key-value reuse from an active request-sending client~\cite{spliceleak}. That attacker queries the victim endpoint, reads the time to first token, and targets the input prompt. Our attacker is a non-querying co-tenant on a separate GPU that sends the victim nothing, reads a self-timed host-memory probe, and observes generation-phase device-to-host traffic, so the two works differ in both threat model and target secret.

\parhead{Covert-Channel Capacity and Defenses.} Robust coding lifts a contention covert channel to carry interactive traffic~\cite{maurice2017hello}, and the standard defenses partition or randomize the shared resource through cache partitioning~\cite{liu2016catalyst}, new cache designs~\cite{wang2007new}, and memory placement~\cite{kim2012stealthmem}. Our placement defense follows this spirit, since separating tenants onto distinct memory controllers removes the shared substrate, and our footprint-flattening defense attacks the signal itself. Relative to this line of work, our contribution is the setting and the signal. We study a contention channel between two physically separate, non-NVLinked GPUs on separate IOMMU units that the DMA footprint of the victim drives, that an unprivileged, performance-counter-free, self-timed GPU kernel mounts, and that passively detects a real co-resident language model instead of recovering a cryptographic secret from a controlled sender.

\section{Limitations and Generalization}
\label{sec:limits}
We report how far the channel generalizes, where it weakens or fails, and what evidence backs each load-bearing claim. The paragraphs below cover the platforms, the replication status, the victim regimes that leak, and the constraints on the attacks.

\parhead{Cross-Platform Generalization.} The channel reproduces on three platforms spanning two GPU generations. On the Rental (Blackwell), the device-to-host channel reproduces with a probe-latency shift of 15 to 19~ns, positive in 93\% to 100\% of paired trials, while host-to-device transfers stay inconsistent, so direction asymmetry holds across generations. Footprint sensitivity also reproduces, though its shape is platform-dependent: monotonic with a knee on the Workstation, a peaked band on the Rental. On the 2-socket server, a device-order-verified grid confirms that the signal depends jointly on footprint and bandwidth, and the cross-NUMA test of Section~\ref{sec:reach} localizes the contended resource to the memory controller. We could not repeat the \texttt{iommu=off} control on the Rental or the 2-socket server, which we do not administer, so it remains unconfirmed whether the same IOMMU page-walk contention underlies the reproduced signal on those machines. We do not claim that the channel works on any GPU, vendor, or NIC. Nevertheless, the underlying mechanism, IOMMU page-walk contention in shared host DRAM, is architectural rather than implementation-specific: any platform that translates device DMA through host-resident page tables shares the structural prerequisite, so the channel is likely to generalize beyond the three platforms tested here.

\parhead{Replication Status.} The core workstation channel is confirmed across five independent runs, and the \texttt{iommu=off} control is replicated as an on/off/on cycle. Cross-NUMA controller locality rests on a device-order-verified run with a symmetric cross-socket control and an earlier agreeing measurement on the Rental. The leakage phase diagram rests on one clean run. Independent replication and testing on additional vendors remain future work. The \texttt{iommu=off} control and the matched-bandwidth distinct-page contrast point to IOMMU page-walk contention in shared host DRAM but do not by themselves separate it from other DMA-path changes. An \texttt{iommu=pt} passthrough control would further isolate them. Despite these open edges, the load-bearing claims, channel existence, footprint sensitivity, direction asymmetry, IOMMU dependence, and memory-controller locality, are each supported by at least one replicated or multi-condition control, and no result relies on a single unreplicated trial.

\parhead{Victim Regime and Cost.} Exploitability tracks the host-transfer regime more than the serving engine. A production vLLM+LMCache key-value offload does not leak across four configurations, consistent with its bursty, low-average device-to-host traffic, while the SGLang CPU weight-offload victim, whose traffic is sustained, leaks at a Spearman correlation of 0.996. Both inferences also require the shared host memory subsystem to be under load, and the inference does not succeed on an otherwise-idle host. The probe itself imposes no measurable load: a co-resident SGLang victim generates at an identical 3.06 tokens per second whether or not the probe runs. The drift mitigation of Section~\ref{sec:robust} cancels drift in a controlled comparison but does not establish an online attack that uses only attacker-controlled operations. The regime restriction does not diminish the security relevance of the channel: sustained device-to-host traffic is the normal operating mode of CPU weight-offload deployments, tensor-parallel collective spills, and large-batch checkpoint writes, all of which are common in cost-sensitive multi-tenant GPU clusters where the threat model applies.

\parhead{Attack-Level Constraints.} Three constraints bound the demonstrated attacks. First, output-length recovery requires successful request segmentation from a noisy probe trace. Segmentation fails on roughly one third of fresh victim processes (Section~\ref{sec:llm}), so the 0.99 Spearman correlation is conditional on clean segmentation. Second, refusal inference is model-dependent: it requires systematically shorter refusals, which Qwen2.5 produces but Mistral-7B does not. Third, the production vLLM+LMCache stack does not leak because its traffic is bursty with low time-averaged occupancy. The channel's operating regime is sustained device-to-host traffic above a floor of about 2~GB/s. These constraints limit the current attacks but not the channel itself. The channel exists at an AUC of 1.000 under controlled conditions regardless of segmentation quality or model behavior, so the constraints belong to the attack pipeline and leave the underlying isolation gap intact. A stronger segmentation algorithm, a different behavioral feature, or a shift in deployment patterns toward heavier host offload could each relax one of these constraints without requiring a new channel primitive.

\section{Conclusion}
We identify a footprint-sensitive timing channel between physically separate GPUs without NVLink that share a host memory subsystem. The channel requires neither privileges nor hardware counters, crosses process, container, and user boundaries, and reproduces across two GPU generations. We demonstrate covert communication at tens of bits per second, output-length inference with a Spearman correlation of 0.99, and refusal inference with balanced accuracy up to 0.955. Exploitation depends on reliable request segmentation, which fails for roughly one third of fresh victim processes, and sustained device-to-host traffic above a bandwidth floor. Refusal inference additionally requires systematically shorter refusals. Placing tenants on separate memory controllers nearly eliminates the signal, while footprint flattening with a 256~MiB pad reduces adaptive inference to chance. The underlying page-walk contention highlights a broader lesson: isolating accelerators does not isolate the host infrastructure they share.

\bibliographystyle{IEEEtran}
\bibliography{references}

\appendices
\section{Ethics Considerations and Vulnerability Disclosure}
\label{app:ethics}

\parhead{Experimental Scope.} This work studies a hardware timing channel between tenants on a shared multi-GPU host. All experiments ran on hardware that we own or rented under our own accounts, with no third-party tenant present, so no data or workload of any other user was observed. The victim workloads are our own processes and language-model servers running our own inputs, the covert-channel sender is a colluding process that we control, and we never probed a workload of another party. We identify no human-subjects concerns and collected no personal data.

\parhead{Potential Harm and Mitigations.} The channel already exists in deployed commodity hardware, and disabling hardware performance counters, the common hardening advice, does not close it, since our primitive is self-timed. We judge that publishing strengthens defenses more than it aids attackers, because anyone with local GPU access can measure the channel and the paper supplies the placement and footprint mitigations that reduce it.

\parhead{Vulnerability Disclosure.} We plan to disclose our findings to the relevant hardware and cloud vendors and provide reproduction instructions, evidence from the tested configurations, and the evaluated mitigations. This manuscript does not report any completed disclosure.
\end{document}